\documentclass[aps,prb,showpacs,twocolumn,superscriptaddress]{revtex4}
\usepackage{graphicx}
\usepackage{amsmath}

\begin{document}

\title{Magnetic exchange in $\alpha $--iron from the ab initio calculations
in the paramagnetic phase}
\author{P. A. Igoshev}
\affiliation{Institute of Metal Physics, Russian Academy of Sciences, 620990
Ekaterinburg, Russia}
\affiliation{Ural Federal University, 620002 Ekaterinburg, Russia}
\author{A. V. Efremov}
\affiliation{Institute of Metal Physics, Russian Academy of Sciences, 620990
Ekaterinburg, Russia}
\author{A. A. Katanin}
\affiliation{Institute of Metal Physics, Russian Academy of Sciences, 620990
Ekaterinburg, Russia}
\affiliation{Ural Federal University, 620002 Ekaterinburg, Russia}
\date{\today }

\begin{abstract}
Applying the local density approximation (LDA) and dynamical mean field
theory (DMFT) to paramagnetic $\alpha $--iron, we revisit a problem of
theoretical description of its magnetic properties. The analysis of local
magnetic susceptibility shows that at sufficiently low temperatures $T<1500$%
~K, both, $e_{g}$ and $t_{2g}$ states equally contribute to the formation of
the effective magnetic moment with spin $S=1$. The self-energy of $t_{2g}$
states shows sizable deviations from Fermi-liquid form, which accompanies
earlier found non-quasiparticle form of $e_{g}$ states. By considering the
non-uniform magnetic susceptibility we find that the non-quasiparticle form
of $e_{g}$ states 
is crucial for obtaining ferromagnetic instability in $\alpha $-iron. The
main contribution to the exchange interaction, renormalized by the effects
of electron interaction, comes from the hybridization between $t_{2g}$ and $%
e_{g}$ states. We furthermore suggest the effective spin-fermion model for $%
\alpha $-iron, which allows us to estimate the exchange interaction from
paramagnetic phase, which is in agreement with previous calculations in the
ordered state within the LDA approaches.
\end{abstract}

\pacs{Valid PACS appear here}
\maketitle



Elemental iron in its low-temperature body-centered cubic (bcc) phase, which
is stable below approximately 1200~K, provides unique example of itinerant
magnetic $d$-electron systems, where formation of well-defined local
magnetic moments can be expected. Indeed, the Rhodes-Wolfarth ratio $p_{%
\mathrm{C}}/p_{\mathrm{S}}$ for this substance is very close to one, which
is characteristic feature of systems, containing (almost) localized $d$%
--electrons ($p_{\mathrm{C}}$ corresponds to the magnetic moment, extracted
from the Curie--Weiss law for magnetic susceptibility in the paramagnetic
phase $\chi =(g\mu _{\mathrm{B}})^{2}p_{\mathrm{C}}(p_{\mathrm{C}}+1)/T$,
and $p_{\mathrm{S}}$ is the saturation moment, $g$ is a Lande factor, $T$
denotes temperature). At the same time, the moment $p_{\mathrm{C}}=1.1$ has
a small fractional part, which is natural for the itinerant material.

This poses natural questions: which electrons mainly contribute to the
local-moment spin degrees of freedom of $\alpha $-iron? What is the
appropriate physical model, that describes spin degrees of this substance?
Attempting to answer the former question, Goodenough suggested\cite%
{Goodenough}, that the $e_{g}$ electrons are localized, while $t_{2g}$
electrons are itinerant. This suggestion was later on refined in Ref. %
\onlinecite{IKT1993}, pointing to a possibility, that only some fraction of $%
e_{g}$ electrons, contributing to formation of the peak of the density of
states near the Fermi level, named by the authors as giant van Hove
singularity, is localized. (The intimate relation between peaks of density
of states and electron localization was also previously pointed out in Ref. %
\onlinecite{Vonsovskii_FMM_1993}). On contrary, there were statements made
that $95\%$ of electrons are localized in iron\cite{2band-model}. On the
model side, the thermodynamic properties of $\alpha $-iron were described
within the effective spin $S=1$ Heisenberg model \cite{Neugebauer}, assuming
therefore that the main part of magnetic moment is localized, in agreement
with the above-mentioned Rhodes-Wolfarth arguments.
Use of the effective Heisenberg model was justified from the ab--initio
analysis of spin spiral energies yielding reasonable values of the exchange
integrals \cite{Gornostyrev}.

These considerations however did not take into account strong electronic
correlations in $\alpha $--iron, which important role was emphasized first
in Ref. \onlinecite{Lichtenstein_2001}. Previous calculations \cite%
{Katanin2010,Abrikosov2013} within the local density approximation (LDA),
combined with the dynamical mean-field theory (DMFT) revealed the presence
of non-quasiparticle states formed by $e_{g}$ electrons, which were
considered as a main source of local moment formation in iron, while $t_{2g}$
states were assumed to be itinerant\cite{Katanin2010}. At the same time,
magnetic properties of the same $t_{2g}$ states also show some features of
local-moment behavior. In particular, the temperature dependence of inverse
local spin susceptibility, which was calculated previously\cite{Katanin2010}
only at $T>1000$~K because of the limitations of Hirsch-Fye method, is
approximately linear, including the contribution of $t_{2g}$ states; the
real part of $t_{2g}$ contribution to dynamic local magnetic susceptibility
has a peak at low frequencies, reflecting a possibility of partial local
moment formation by $t_{2g}$ states.

Studying this possibility requires investigation of electronic and magnetic
properties at low temperatures, since the energy scale for partially formed
local $t_{2g}$ moments can be smaller than for $e_{g}$ states. Although real
substance orders ferromagnetically at low temperatures, in the present paper
(as in Ref. \onlinecite{Katanin2010}) we perform analysis of local
properties of iron in the paramagnetic phase to reveal the mechanism of
local moment formation. Furthermore, we study non-local magnetic
susceptibility in the low temperature range $T>250$ $\mathrm{K}$\textrm{,}
which allows us to analyze the mechanism of magnetic exchange. To this end
we use the state--of--art dynamical mean--field theory (DMFT) calculation
with continous time quantum Monte-Carlo (CT--QMC)~solver\cite{WernerQMC},
combined with the ab-initio local density approximation (LDA). From our
low-temperature analysis we argue, that $t_{2g}$ electrons are almost
equally contribute to the effective local magnetic moment, as the $e_{g}$
electrons, and play crucial role in the mechanism of magnetic exchange in
iron. In particular, the most important contribution to the exchange
integrals comes from the hybridization of $t_{2g}$ and $e_{g}$ states, which
yields \textit{nearest-neighbour }magnetic exchange interaction, which
agrees well with the experimental data.

We perform the ab initio band--structure calculations in LDA approximation
within tight--binding--linear muffin--tin orbital--atomic spheres
approximation framework; the von Barth-Hedin local exchange-correlation
potential \cite{Barth_Hedin_1972} was used. Primitive reciprocal translation
vectors were discretized into 12 points along each direction which leads to
72 \textbf{k}--points in irreducible part of the Brillouin zone. For DMFT
(CT-QMC) calculations, we use the Hamiltonian of Hubbard type with the
kinetic term containing all $s$--$p$--$d$ states, being extracted from the
LDA solution, and the interaction part with density-density contributions
for $d$ electrons only. The Coulomb interaction parameter value $U=2.3$~eV
and the Hund's parameter $I=0.9$~eV used in our work are the same as in
earlier LDA+DMFT calculations \cite%
{Lichtenstein_2001,Katanin2010,Igoshev2013}. To treat a problem of formation
of local moments we consider paramagnetic phase, which is achieved by
assuming spin-independent density of states, local self-energy and bath
green function. For the purpose of extracting corresponding exchange
parameters, we take in LDA part physical value of the lattice parameter $%
a=2.8664$~\AA , corresponding to ferromagnetic state at room temperature.%

\begin{figure}[tbp]
\includegraphics[width=0.47\textwidth]{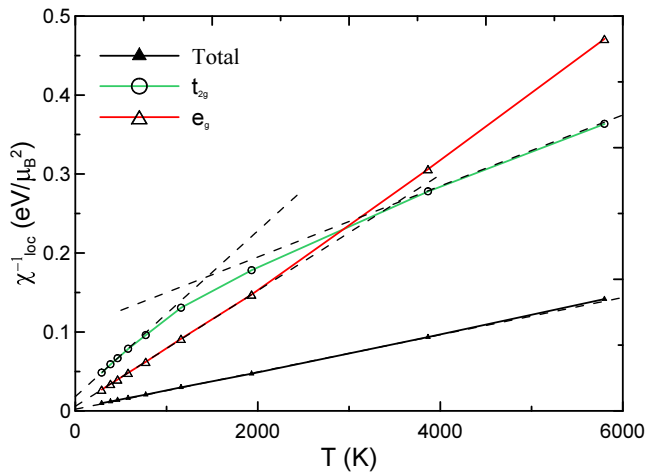}
\caption{(Color online) Temperature dependence of inverse local magnetic
susceptibility, and the corresponding $e_{g}$ and $t_{2g}$ orbital
contributions. Dashed lines show linear behavior in different temperature
intervals. }
\label{fig:inverse_chi}
\end{figure}

We consider first the results for the orbital--resolved
temperature--dependent local static spin susceptibility $\chi _{\text{loc,}%
mn}=4\mu _{B}^{2}\int\nolimits_{0}^{\beta }\langle s_{i,m}^{z}(\tau
)s_{i,n}^{z}(0)\rangle \,d\tau $, where $s_{i,m}^{z}$ is the $z$--projection
of the spin of $d$--electrons, belonging to the orbitals $m=t_{2g},e_{g}$ at
a given lattice site $i$, see Fig. \ref{fig:inverse_chi} (for completeness,
we also show the total susceptibility $\chi _{\text{loc}}=\sum\nolimits_{mn}%
\chi _{\text{loc,}mn}$ which also includes the off-diagonal $t_{2g}$-$e_{g}$
contribution). The temperature \ dependence of the static inverse local
susceptibility is linear (as was also observed in previous studies \cite%
{Lichtenstein_2001,Katanin2010,Abrikosov2013,Igoshev2013}), however being
resolved with respect to orbital contributions (see Fig. \ref%
{fig:inverse_chi}) it appears to manifest very different nature of $e_{g}$
and $t_{2g}$ moments. The inverse $e_{g}$ orbital contribution behaves
approximately linearly with $T$ in a broad temperature range\cite%
{Katanin2010,Abrikosov2013}. At the same time, analyzing low-temperature
behaviour, we find that $\chi _{\text{loc,}t_{2g}\text{-}t_{2g}}^{-1}$
demonstrates a crossover at $T^{\ast }\sim 1500$~K between two linear
dependences with the low--temperature part having higher slope (i. e.
smaller effective moment). Note that this feature was not obtained in
previous study \cite{Katanin2010} because of considering only temperature
range $T>1000$ K. The scale $T^{\ast }$ corresponds to the crossover to
non-Fermi-liquid behavior of $t_{2g}$ states, see below.

\begin{figure}[tbp]
\includegraphics[width=0.47\textwidth]{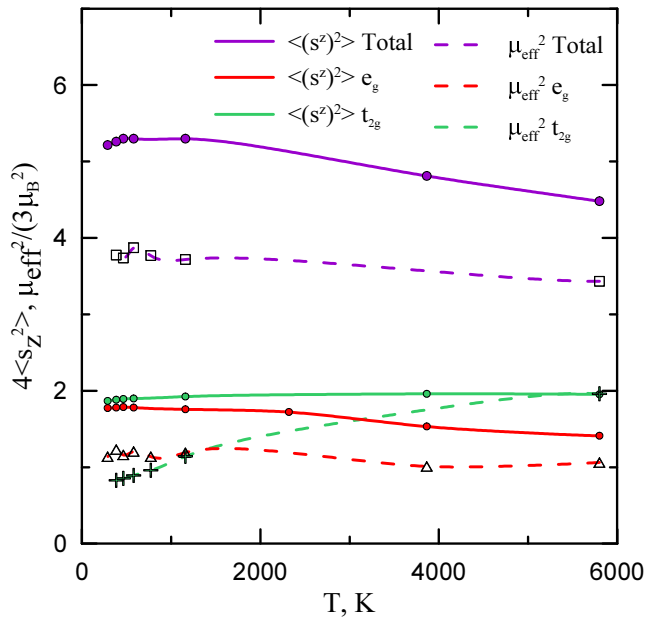}
\caption{(Color online) The temperature dependence of the effective magnetic
moment and instantaneous average $\langle (s^{z})^{2}\rangle $ and $\protect%
\mu _{\mathrm{eff}}^{2}$ in $\protect\alpha $--iron, extracted from the
temperature dependence of local susceptibility, together with the
contribution of the $e_{g}$ and $t_{2g}$ orbitals}
\label{fig:moms_vs_T}
\end{figure}

To get further insight into the local magnetic properties of $\alpha $-iron,
we consider the temperature dependence of the effective magnetic moment $\mu
_{m,\mathrm{eff}}^{2}=3/(d\chi _{\text{loc,}mm}^{-1}/dT)$ and the
instantaneous average $\langle (s_{i,m}^{z})^{2}\rangle ,$ corresponding to
different orbital states, see Fig. \ref{fig:moms_vs_T}. We find, that for $%
e_{g}$ electrons both moments saturate at temperatures $T<1500$~K and remain
approximately constant up to sufficiently low temperatures. Comparing the
value of the square of the moment $\mu _{e_{g},\mathrm{eff}}^{2}/(3\mu
_{B}^{2})=1.2$, extracted from the Curie-Weiss law for local susceptibility,
and the instantaneous average $4\langle (s_{i,e_{g}}^{z})^{2}\rangle =1.8$
with the corresponding filling $n_{e_{g}}\simeq 2.6$, we find that the major
part of $e_{g}$ electrons determine the instantaneous average $\langle
(s_{i,e_{g}}^{z})^{2}\rangle $, and at least half of them contribute to the
sufficiently long-living (on the scale of $1/T$) local moments. At the same
time, for $t_{2g}$ electronic states the abovementioned crossover between
the high-temperature value $\mu _{t_{2g},\mathrm{eff}}^{2}/(3\mu
_{B}^{2})\approx 1.95$ and the low temperature value $\mu _{t_{2g},\mathrm{%
eff}}^{2}/(3\mu _{B}^{2})\simeq 0.82$ is present, which, comparing to $%
n_{t_{2g}}\simeq 4.4,$ shows that at least $20\%$ of $t_{2g}$ electrons
participate in the effective local moment formation at low temperatures.
Yet, the corresponding low-temperature effective moments $\mu _{e_{g},%
\mathrm{eff}}^{2}$ and $\mu _{t_{2g},\mathrm{eff}}^{2}$ are comparable (each
of them is approximately $3\mu _{B}^{2},$ corresponding to the effective
spin $s\simeq 1/2$), showing important role of $t_{2g}$ electrons in the
formation of the total spin $S=1$ state.

\begin{figure}[h]
\caption{ (Color online) Frequency dependence of $\mathrm{Im}\Sigma _{t_{%
\mathrm{2g}}}(\mathrm{i}\protect\nu )$. Black solid lines illustrate the
results of calculations at temperatures (from top to bottom) $%
T=1/5,1/10,1/20,1/30,1/40$ eV. The green dot--dashed line present fits to
the the Fermi-liquid dependence in the range $\protect\nu _{n}<1$~eV, while
blue dashed lines present fits to the non-Fermi-liquid dependence (see
text). Dots denote Matsubara frequencies $\protect\nu _{n}=\protect\pi %
T(2n+1)$.}
\label{fig:sigma}\includegraphics[width=0.44\textwidth]{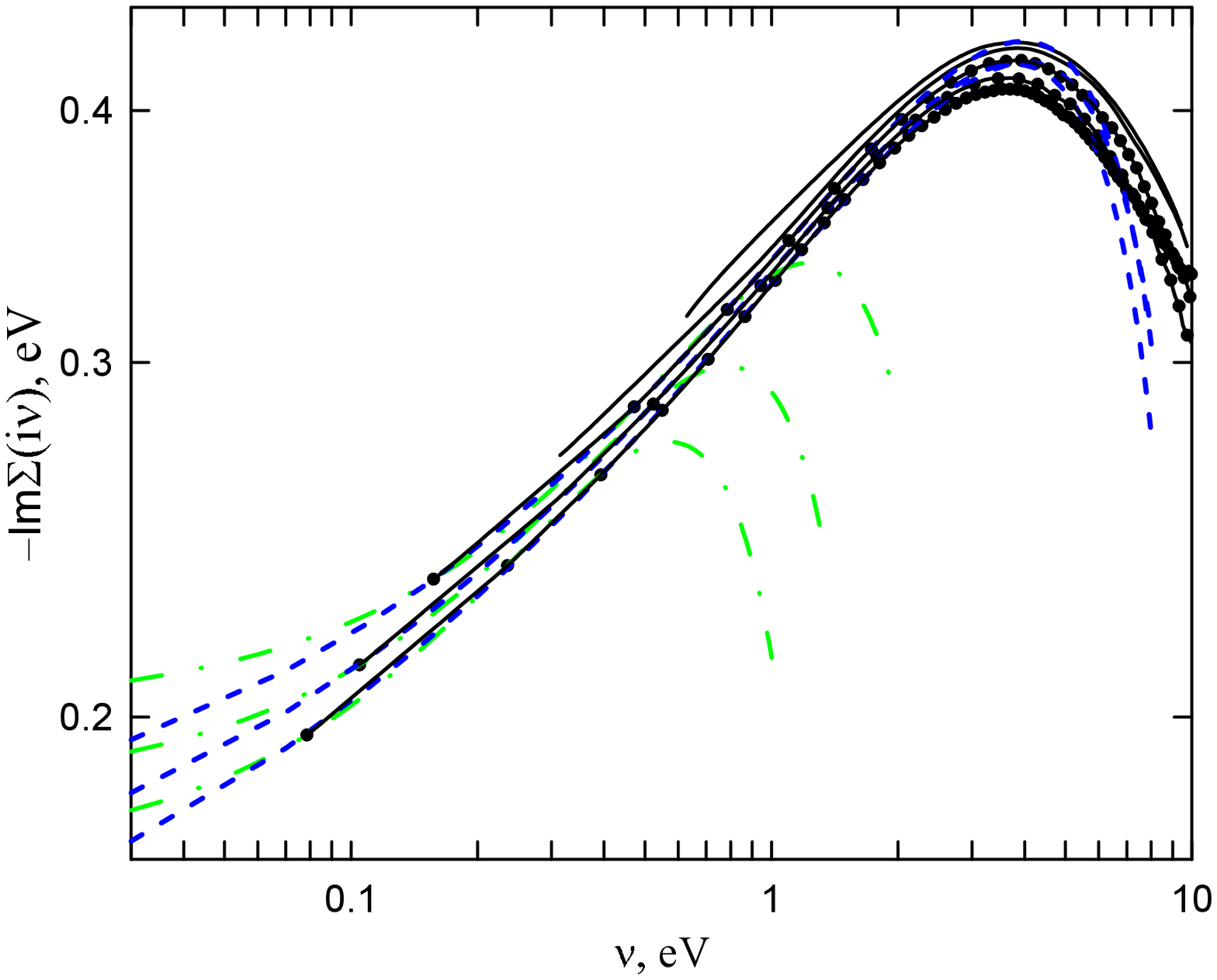}
\end{figure}

Although the self-energy calculations \cite{Katanin2010,Abrikosov2013} yield
quasiparticle-like form of $t_{2g}$ electron self-energy, the low-frequency
and low-temperature dependence of self-energy shows pronounced deviations
from the Fermi-liquid behavior, see Fig. \ref{fig:sigma}. To analyse the
frequency dependence of the self-energy on imaginary frequency axis, we fit
the obtained results by the Fermi-liquid dependence $-\mathrm{Im}\Sigma
(i\nu )=\Gamma (T)+[Z^{-1}(T)-1]\nu +\sigma (T)\nu ^{2},$ where $\Gamma (T)$
is the damping of electrons at the Fermi level, $Z(T)$ is the
temperature-dependent quasiparticle residue. Alternatively, we consider the
fit $-\mathrm{Im}\Sigma (i\nu )=\Gamma _{1}(T)+\beta _{1}(T)\nu ^{\alpha
}+\sigma _{1}(T)\nu ^{2}$ with some exponent $\alpha <1$. The latter
dependence corresponds to the non-Fermi-liquid behavior of $t_{2g}$
electrons. The obtained results are presented in the Table.
\begin{equation*}
\begin{tabular}{||l||l|l|l||l|l|l|l||}
\hline\hline
$\beta =1/T$ & $\Gamma $ & $Z^{-1}-1$ & $\sigma $ & $\Gamma _{1}$ & $\beta
_{1}$ & $\sigma _{1}$ & $\alpha $ \\ \hline\hline
20 & 0.20 & \multicolumn{1}{|c|}{0.22} & -0.09 & 0.17 & 0.18 & -0.006 & 0.51
\\
30 & 0.18 & \multicolumn{1}{|c|}{0.29} & -0.19 & 0.15 & 0.19 & -0.005 & 0.48
\\
40 & 0.17 & \multicolumn{1}{|c|}{0.37} & -0.32 & 0.13 & 0.20 & -0.005 & 0.44
\\ \hline\hline
\end{tabular}%
\end{equation*}

The linear-quadratic fits are applicable only at $\nu <1$ eV; at
sufficiently small $\nu $ they also do not fit the obtained results well. We
find that the spectral weight $Z(T)$ pronouncely decreases with decrease of
temperature, and the coeffitient $\Gamma (T)$ obviously does not obey the
Fermi-liquid dependence $\Gamma (T)\propto T^{2}$. These observations show
that sizable deviations from Fermi-liquid picture can be expected.

The power-law fits yield much better agreement in a broad range of
frequencies $\nu <5$ eV, describing at the same time correctly the
low-frequency behavior. The coefficients $\beta _{1},$ $\sigma _{1}$ of
these fits show very weak temperature dependence (the contribution $\sigma
_{1}$ is almost negligible), while the damping $\Gamma _{1}(T)$ and the
exponent $\alpha $ slightly decrease with temperature, being related by $%
\Gamma _{1}(T)\sim T^{\alpha }$. These observations imply that $t_{2g}$
electronic subsystem is better described by non-Fermi liquid behavior at low
temperatures, which reflects its participation in the formation of local
moments in $\alpha $-iron. Remarkably, consideration of the three-band model
in Ref. \onlinecite{Werner} showed similar dependence of the self-energy $%
\Sigma \sim \nu ^{1/2}$ due to Hund exchange interaction, which allows to
attribute the $t_{2g}$ subsystem in iron as close to the "spin
freesing\textquotedblright\ transition, accroding to the terminology of Ref. %
\onlinecite{Werner}.

To get further insight into the formation of effective local moments and
extract corresponding exchange integrals,
we calculate the momentum dependence of particle-hole bubble $\chi _{\mathbf{%
q}}^{\text{\textrm{0}}\mathrm{,}mn}=-(2\mu _{B}^{2}/\beta )\sum_{l,{\mathbf{%
k,}}\widetilde{m}\in m,\widetilde{n}\in n}\mathcal{G}_{\mathbf{k},\widetilde{%
m}\widetilde{n}}(\mathrm{i}\nu _{l})\mathcal{G}_{\mathbf{k}+\mathbf{q},%
\widetilde{n}\widetilde{m}}(\mathrm{i}\nu _{l})$, which is obtained using
paramagnetic LDA and LDA+DMFT electronic spectrum [$\mathcal{G}_{\mathbf{k},%
\widetilde{m}\widetilde{n}}(\mathrm{i}\nu _{l})$ is the corresponding
electronic Green function for the transition from the orbital state $%
\widetilde{m}$ to $\widetilde{n}$, $\nu _{l}$ is a fermionic Matsubara
frequency; for more details on the calculation procedure see Ref. %
\onlinecite{Igoshev2013}].
\begin{figure}[h]
\includegraphics[width=0.44\textwidth]{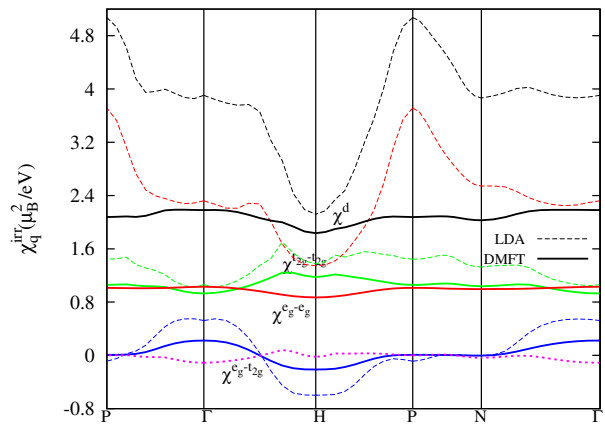}
\caption{ (Color online) Orbital-resolved momentum dependence of $\protect%
\chi _{\mathbf{q}}^{\mathrm{0},mn}$ at $T=290$~K calculated in high symmetry
directions of the Brillouin zone. The contributions $\protect\chi _{\mathbf{q%
}}^{\mathrm{0},d}$, $\protect\chi _{\mathbf{q}}^{\mathrm{0},t_{2g}-t_{2g}}$,
$\protect\chi _{\mathbf{q}}^{\mathrm{0},e_{g}-e_{g}}$ , and the
hybridization part $\protect\chi _{\mathbf{q}}^{\mathrm{0},e_{g}-t_{2g}}$
are shown by black, red, green and blue lines, respectively. Solid (dashed)
lines correspond to LDA+DMFT (LDA) results. The LDA+DMFT estimate for $%
J_{q}^{(1)}(\protect\mu _{B}/I)^{2}$ is shown by magenta short--dashed line.
}
\label{fig:chi:lda_vs_dmft}
\end{figure}
The results for LDA and LDA+DMFT approaches at $T=290$~K are presented in
the Fig.~\ref{fig:chi:lda_vs_dmft} (we find that the LDA+DMFT results are
almost temperature-independent at low $T$). For the bubble, calculated using
purely LDA spectrum (i.e. with the assumption that all electrons are
itinerant), the maximum of $\chi _{\mathbf{q}}^{\mathrm{0}}$ is located at
the point $\mathbf{q}=\mathbf{q}_{\mathrm{P}}=(\pi ,\pi ,\pi )/a$, while the
ferromagnetic instability in $\alpha $-iron requires maximum of $\chi _{%
\mathbf{q}}^{\mathrm{0}}$ at $\mathbf{q}=0$ and low $T,$ if one neglects the
non-local vertex corrections. One can observe, that the main contribution to
this \textquotedblleft incorrect\textquotedblright\ behavior of the bubble
originates from the $e_{g}$ electron part, $\chi _{\mathbf{q}}^{\mathrm{0}%
,e_{g}-e_{g}}$. Both $\chi _{\mathbf{q}}^{\mathrm{0},e_{g}-e_{g}}$ and $\chi
_{\mathbf{q}}^{\mathrm{0},e_{g}-t_{2g}}$ contributions are however strongly
influenced 
by the account of the local self--energy corrections to the Green's function
in DMFT approach, which correspond physically to account of partial
localization of $d$-electrons. These corrections mainly change $\chi _{%
\mathbf{q}}^{\mathrm{0},e_{g}\text{-}e_{g}}$ and shift the maximum of $\chi
_{\mathbf{q}}^{\mathrm{0}}$ to $\Gamma $ point ($\mathbf{q}=0$). Note that
within LDA+DMFT, intra--orbital contributions to $\chi _{\mathbf{q}}^{%
\mathrm{0},e_{g}-e_{g}}$ and $\chi _{\mathbf{q}}^{\mathrm{0},t_{2g}-t_{2g}}$
are only weakly momentum-dependent; they also behave similarly, varying
\textquotedblleft counter--phase\textquotedblright . According to the
general ideas of spin-fluctuation theory \cite{Moriya}, this weak momentum
dependence can be ascribed to the formation of the effective moments from $%
e_{g}$ and $t_{2g}$ states. In agreement with the abovediscussed
consideration, the $\chi _{\mathbf{q}}^{\mathrm{0},e_{g}\text{-}e_{g}}$
contribution has even weaker dispersion than the $\chi _{\mathbf{q}}^{%
\mathrm{0},t_{2g}\text{-}t_{2g}}$ part. At the same time, strongly
dispersive $\chi _{\mathbf{q}}^{\mathrm{0},e_{g}-t_{2g}}$ contribution,
which is assumed to correspond to the (remaining) itinerant degrees of
freedom, provides the maximum of the resulting $\chi _{\mathbf{q}}^{\mathrm{0%
}}$ at $\mathbf{q}=0$ and appears to be the main source of the stability of
the ferromagnetic ordering in iron within LDA+DMFT approximation.

The obtained results do not change qualitatively for the other choice
Hubbard interactions (as we have verified for $U=4.0$ and $I=1.0$~eV), see
Supplementary Material \cite{Suppl}.

To see the quantitative implications of the described physical picture, we
consider the effective spin-fermion model
\begin{align}
\mathcal{S}& =\frac{1}{2}\sum_{i,\omega _{n}}\chi _{S}^{-1}(\mathbf{q},%
\mathrm{i}\omega _{n})\mathbf{S}_{i}(\mathrm{i}\omega _{n})\mathbf{S}_{j}(-%
\mathrm{i}\omega _{n})e^{i\mathbf{q(R}_{i}-\mathbf{R}_{j}\mathbf{)}}
\label{SF} \\
& +2I\sum_{i,\omega _{n}}\mathbf{S}_{i}(\mathrm{i}\omega _{n})\mathbf{s}%
_{i}(-\mathrm{i}\omega _{n})  \notag \\
& +\sum_{\nu _{n}\sigma ll^{\prime }}c_{l\sigma }^{\dag }(\mathrm{i}\nu _{n})%
\left[ \mathrm{i}\nu _{n}\delta _{ll^{\prime }}+H_{ll^{\prime }}+\Sigma
_{ll^{\prime }}(\mathrm{i}\nu _{n})\right] c_{l^{\prime }\sigma }(\mathrm{i}%
\nu _{n})  \notag
\end{align}%
($\omega _{n}$ is a bosonic Matsubara frequency, $l,l^{\prime }$ combines
site and orbital indices), describing interaction of itinerant electrons
with (almost) \textit{local} spin fluctuations (in contrast to critical spin
fluctuation in cuprates \cite{Schmalian}), see also Ref. \cite{DGA1}. We
assume here that the Coulomb and Hund's interaction acting within $e_{g}$
and $t_{2g}$ orbitals results in a formation of some common local moment
(field $\mathbf{S}$), while the remaining itinerant degrees of freedom are
described by the field $\mathbf{s}_{i}=\mathbf{s}_{i}^{e_{g}}+\mathbf{s}%
_{i}^{t_{2g}},$ formed from the Grassmann variables $c_{l^{\prime }\sigma };$
$H_{ll^{\prime }}$ and $\Sigma _{ll^{\prime }}$ are the Hamiltonian and
local self-energy corrections to the LDA spectrum (the latter is assumed to
be local and therefore diagonal with respect to orbital indices). The
interaction between the two subsystems (localized and itinerant), which are
formed from the $d$--electronic states, is driven by Hund's constant
coupling $I$.

Considering the renormalization of the propagator $\chi _{S}$ by the
corresponding boson self--energy corrections, we obtain for the non-uniform
susceptibility (see Supplementary Material \cite{Suppl})
\begin{equation}
\chi ^{-1}(\mathbf{q},\mathrm{i}\omega _{n})=\chi _{\mathrm{loc}}^{-1}(%
\mathrm{i}\omega _{n})-J_{\mathbf{q}}/(4\mu _{B}^{2}),  \label{chiq}
\end{equation}%
where $\chi _{\mathrm{loc}}(\mathrm{i}\omega _{n})$ is the local spin
susceptibility and $J_{\mathbf{q}}$ is the exchange interaction, which
fulfills $\sum\nolimits_{\mathbf{q}}J_{\mathbf{q}}=0$ (no spin
self-interaction). We find $J_{\mathbf{q}}=J_{\mathbf{q}}^{(1)}+J_{\mathbf{q}%
}^{(2)}$, $J_{\mathbf{q}}^{(1)}=(I/\mu _{B})^{2}\sum_{m}\left[ \chi _{%
\mathbf{q}}^{\mathrm{0,}mm}-\sum\nolimits_{\mathbf{p}}\chi _{\mathbf{p}}^{%
\mathrm{0,}mm}\right] $ is the intra-orbital part, while $J_{\mathbf{q}%
}^{(2)}=2(I/\mu _{B})^{2}\chi _{\mathbf{q}}^{\mathrm{0,}t_{2g}\text{-}e_{g}}$
results from the hybridization of states of different symmetry. The
contribution $J_{\mathbf{q}}^{(1)}$ is approximately twice smaller than $J_{%
\mathbf{q}}^{(2)},$ and therefore the main contribution to the magnetic
exchange comes from the hybridization of $t_{2g}$ and $e_{g}$ states. The
whole momentum dependence of $J_{\mathbf{q}}^{(2)}$ can be well captured by
the nearest--neighbor approximation for effective exchange integrals only, $%
J_{\mathbf{q}}^{(2)}=J_{0}\cos (aq_{x}/2)\cos (aq_{y}/2)\cos (aq_{z}/2),$
while $J_{\mathbf{q}}^{(1)}$ has more complicated momentum dependence.

%

Restricting ourselves by considering the contribution $J_{\mathbf{q}}=J_{%
\mathbf{q}}^{(2)}$, (we assume that the contribution $J_{\mathbf{q}}^{(1)}$
is further suppressed by the non-local and vertex corrections), from Fig. %
\ref{fig:inverse_chi} we find at $T=290$~K the value $J_{\mathbf{q}=0}=0.18$%
~eV. This value, as well as the momentum dependence of $J_{\mathbf{q}}^{(2)}$
agrees well with the result of S.V. Okatov et al. \cite{Gornostyrev}. %
The obtained results together with $\mu _{\mathrm{eff}}^{2}=11.4\mu _{%
\mathrm{B}}^{2}$ (see Fig. 2) provide an estimate for the Curie temperature
(we assume $T_{\mathrm{C}}\gg \theta $), which can be obtained from the
divergence of $\chi ^{-1}(\mathbf{q},\mathrm{0})$:
\begin{equation}
T_{\mathrm{C}}=\frac{\mu _{\mathrm{eff}}^{2}}{4\mu _{\mathrm{B}}^{2}}\frac{%
J_{0}}{3}=0.17\text{ eV}
\end{equation}%
and appears comparable with the result of full DMFT calculation, and
therefore shows that the above model is adequate for describing magnetic
properties of the full $5$-band Hubbard model. (Note that the overestimation
of $T_{\mathrm{C}}$ in DMFT approach in comparison with the experimental
data is due to density-density approximation for the Coulomb interaction\cite%
{Anisimov} and (to minor extent) due to presence of non-local fluctuations,
not accounted by DMFT).

Neglecting longitudinal fluctuations of field $\mathbf{S}$ we can map the
model (\ref{SF}) to an effective $S=1$ Heisenberg model $\mathcal{H}_{%
\mathrm{H}}=(1/2)\sum_{ij}J_{ij}\mathbf{S}_{i}\mathbf{S_{j}}$ \ to estimate
the spin--wave spectrum:
\begin{equation}
\omega _{\mathbf{q}}=S(J_{0}-J_{\mathbf{q}})=S(I/\mu _{B})^{2}(\chi _{0}^{%
\mathrm{0},e_{g}-t_{2g}}-\chi _{\mathbf{q}}^{\mathrm{0},e_{g}-t_{2g}}).
\label{eq:spin-wave_spectrum}
\end{equation}%
We obtain the corresponding spin stiffness $D=\lim_{q\rightarrow 0}(\omega _{%
\mathbf{q}}/q^{2})=290\,\text{meV}\cdot $\AA $^{2}$ in a good agreement with
the experimental data $D=280\,\text{meV}\cdot $\AA $^{2}$ (Ref.~%
\onlinecite{Mook1973}). 

In conclusion, we have considered the problem of the description of
effective local moments in $\alpha $-iron based on the electronic spectrum
in paramagnetic phase within LDA+DMFT approximation. We find that local
moments are formed by both $e_{g}$ and $t_{2g}$ orbital states, each of them
contributing a half of the total moment $S=1.$ For $t_{2g}$ electronic
states we find pronounced features of non-Fermi-liquid behavior, which
accompanies earlier observed non-quasiparticle form of $e_{g}$ states. The
local moment and itinerant states interact with itinerant states via Hund
interaction, yielding magnetic exchange between the local-moment states via
the effective RKKY--type mechanism. The obtained exchange integrals are well
captured by the LDA+DMFT approach. The main origin of the intersite
interaction of these moments is attributed to the $e_{g}$-$t_{2g}$
hybridization, which yields magnetic exchange, dominating on the
nearest-neighbour sites. Contrary to the previous studies\cite%
{Lichtenstein1987,Gornostyrev}, we do not however assume some magnetic
ordering for the electronic system.

We also emphasize that non--local self--energy corrections, as well as
vertex corrections, missed in our investigation, can make the described
physical picture more precise. In particular, non-local effects allow for
the non--zero non-diagonal $e_{g}$--$t_{2g}$ self-energy matrix elements and
therefore possibly renormalize the strength of exchange interaction, as well
as the self-energy of $t_{2g}$ electronic states. The role of the vertex
corrections, only roughly accounted in the considered approach, also
requires additional study. Therefore further investigation using powerful
theoretical techniques of dynamic vertex approximation \cite{DGA}, dual
fermion\cite{DF}, or other non-local approaches is of certain interest.


The authors are grateful to Yu.~N.~Gornostyrev, A.~V.~Korolev, and K.~Held
for useful discussions. The work of P. A. Igoshev was supported by the
Russian Foundation for Basic Research (Project No. 14-02-31603) and Act 211
Government of the Russian Federation 02.A03.21.0006; A. A. Katanin
acknowledges support of the Program of "Dynasty" foundation. The
calculations were performed using "Uran\textquotedblright\ supercomputer of
IMM UB RAS.

\clearpage \setcounter{equation}{0} \setcounter{figure}{0}

\section*{Supplementary Material for the paper "Magnetic exchange in $%
\protect\alpha $--iron from the ab initio calculations in the paramagnetic
phase" by P. A. Igoshev et al.}

\subsection{Local and non-uniform susceptibilities for $U=4$ eV}

\begin{figure}[h]
\includegraphics[width=0.47\textwidth]{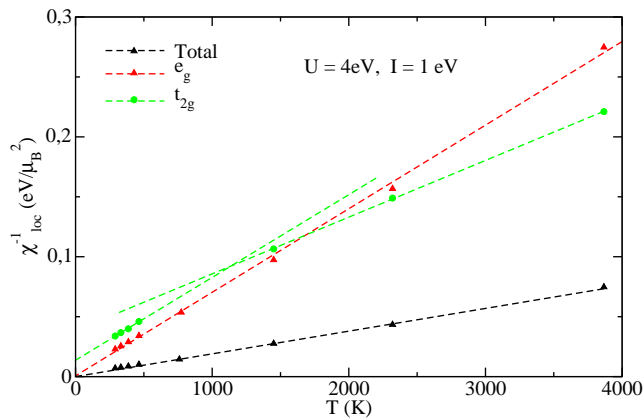}
\caption{(Color online) The same as in Fig. \protect\ref{fig:inverse_chi} of
the main text for $U=4.0 $~and~$I=1.0$~eV.}
\label{fig:chi_loc2}
\end{figure}

\begin{figure}[h]
\includegraphics[width=0.44\textwidth]{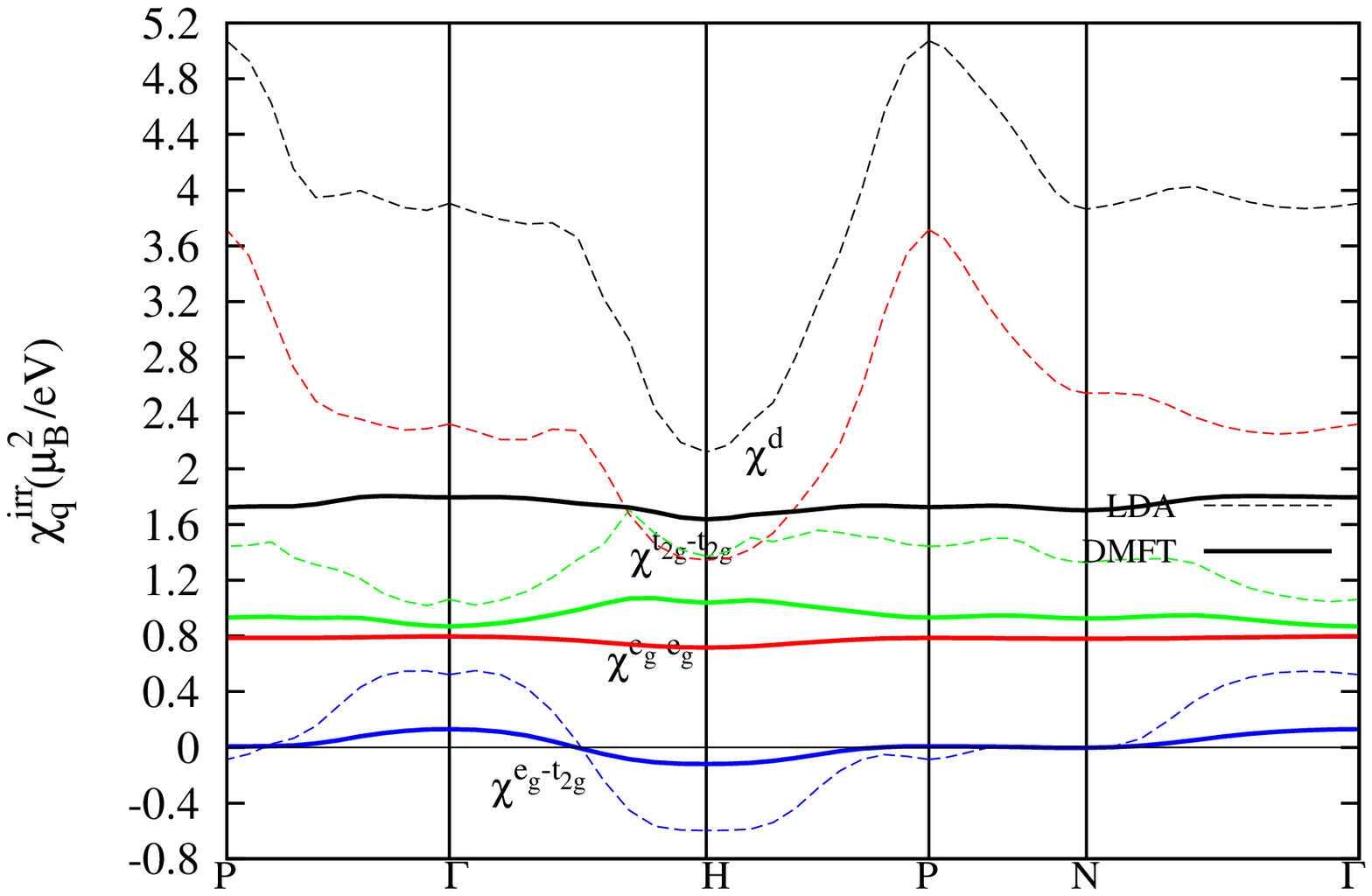}
\caption{ (Color online) The same as in Fig. \protect\ref%
{fig:chi:lda_vs_dmft} of the main text for $U=4.0$~and~$I=1.0$~eV.}
\label{fig:chi_irr_q2}
\end{figure}

We test below the stability of our results to change of model parameter
values: results of the calculations by using the same method as in the main
text but the other choice of parameters ($U=4.0$~and~$I=1.0$~eV), which are
close to those of Ref.~\onlinecite{Pourovskii}. The results for the
temperature dependence of the inverse local magnetic susceptibility are
shown in Fig.~\ref{fig:chi_loc2}. We find the crossover discussed in the
main text at lower $T^{\ast }\sim 1050$~K. The calculation of momentum
dependent irreducible susceptibility yields only the uniform (with respect
to $\mathbf{q}$) renormalization without change of qualitative tendencies
(see Fig.~\ref{fig:chi_irr_q2}, cf.~Fig.~\ref{fig:chi:lda_vs_dmft} of the
main text). We have recalculated exchange interactions from these results
and obtain $J_{\mathbf{q}=0}^{(2)}=0.13$~eV vs 0.18~eV in the main text.
This implies lowering of Curie temperature, which agrees with approximately
renormalization of $T^{\ast }$ by 1.5 times (cf.~Fig.~\ref{fig:inverse_chi}
of the main text). The qualitative conclusions of the paper remain unchanged
for these parameter values. 

\subsection{Calculation of exchange interaction from the spin-fermion model}

To obtain exchange interaction, we first determine the bare propagator of
magnetic degrees of freedom $\chi _{S}(\mathbf{q},\mathrm{i}\omega _{n})$ by
requiring that the dressed propagator of $\mathbf{S}$ field is equal to the
susceptibility of itinerant subsystem. Using the random-phase-type
approximation, which reduces the orbital- and frequency dependence of the
bubble and vertex to the respective single-frequency orbital "averaged"
quantities, $\chi _{\mathbf{q}}^{\mathrm{0}}=\sum_{mn}\chi _{\mathbf{q}}^{%
\mathrm{0,}mn}$ and $\Gamma $, we obtain 
\begin{equation}
\chi _{S}^{-1}(\mathbf{q},\mathrm{i}\omega _{n})=4\mu _{B}^{2}\left( \chi _{%
\mathbf{q}}^{\mathrm{0}}\right) ^{-1}-2\Gamma +(I/\mu _{\mathrm{B}})^{2}\chi
_{\mathbf{q}}^{\mathrm{0}},  \label{chiS}
\end{equation}%
where the last term is added to cancel the corresponding bosonic self-energy
correction from itinerant degrees of freedom to avoid double-counting, cf.
Ref. \cite{OurJETP}. We represent $\chi _{\mathbf{q}}^{\mathrm{0}}=\overline{%
\chi }_{0}+\delta \chi _{\mathbf{q}}^{\mathrm{0}}$ with momentum-independent 
$\overline{\chi }_{0}$; without loss of generality, we can assume $\sum_{%
\mathbf{q}}\delta \chi _{\mathbf{q}}^{\mathrm{0}}=0,$ such that $\overline{%
\chi }_{0}=\sum_{\mathbf{q}}\chi _{\mathbf{q}}^{\mathrm{0}}.$ From the
results of Fig. 4 of the main text it follows that $\delta \chi _{\mathbf{q}%
}^{\mathrm{0}}\ll \overline{\chi }_{0}$. Expanding Eq. (\ref{chiS}) to first
order in $\delta \chi _{\mathbf{q}}^{\mathrm{0}}$, we obtain 
\begin{eqnarray}
\chi _{S}^{-1}(\mathbf{q},\mathrm{i}\omega _{n}) &=&4\mu _{B}^{2}\chi _{%
\mathrm{loc}}^{-1}(\mathrm{i}\omega _{n})+(I/\mu _{\mathrm{B}})^{2}\overline{%
\chi }_{0}  \notag \\
&&+[(I/\mu _{\mathrm{B}})^{2}-4\mu _{B}^{2}\overline{\chi }_{0}^{-2}]\delta
\chi _{\mathbf{q}}^{\mathrm{0}},
\end{eqnarray}%
where $\chi _{\mathrm{loc}}^{-1}(\mathrm{i}\omega _{n})=\overline{\chi }%
_{0}^{-1}-2\Gamma /(4\mu _{B}^{2})$ is the inverse local susceptibility. In
practice, the frequency dependence $\chi _{\mathrm{loc}}(\mathrm{i}\omega
_{n})=\mu _{\mathrm{eff}}^{2}/(3(T+\theta )(1+|\omega _{n}|/\delta ))$ can
be obtained from the dynamic local spin correlation functions, which is
characterized by the temperature-independent moment $\mu _{\mathrm{eff}},$
its damping $\delta \propto T$, and the corresponding Weiss temperature $%
\theta $ 
(see Refs. \onlinecite{Katanin2010,Igoshev2013} of the main text). Since $%
\overline{\chi }_{0}\simeq 2\mu _{B}^{2}/$eV and $I\simeq 1$eV the momentum
dependence is almost cancelled, and we obtain the local bare propagator of
spin degrees of freedom,%
\begin{equation}
\chi _{S}^{-1}(\mathbf{q},\mathrm{i}\omega _{n})\simeq \chi _{S}^{-1}(%
\mathrm{i}\omega _{n})=4\mu _{B}^{2}\chi _{\mathrm{loc}}^{-1}(\mathrm{i}%
\omega _{n})+(I/\mu _{\mathrm{B}})^{2}\overline{\chi }_{0}.
\end{equation}%
Considering the renormalization of the propagator $\chi _{S}$ by the
corresponding boson self--energy corrections (cf. Ref. \cite{OurJETP}), we
obtain for the non-uniform susceptibility 
\begin{equation}
\chi ^{-1}(\mathbf{q},\mathrm{i}\omega _{n})=\frac{1}{4\mu _{B}^{2}}\left[
\chi _{S}^{-1}(\mathbf{q},\mathrm{i}\omega _{n})-\frac{I^{2}}{\mu _{B}^{2}}%
\sum_{mn}\chi _{\mathbf{q}}^{\mathrm{0,}mn}\right] ,
\end{equation}%
which yields Eq. (\ref{chiq}) of the main text (we use also here that by
symmetry $\sum\nolimits_{\mathbf{p}}\chi _{\mathbf{p}}^{\mathrm{0,}t_{2g}%
\text{-}e_{g}}=0$).


\begin{thebibliography}{99}
\bibitem{Goodenough} J. B. Goodenough, Phys. Rev. \textbf{120}, 67 (1960).

\bibitem{IKT1993} V. Yu. Irkhin, M. I. Katsnelson, and A. V. Trefilov, J.
Phys.: Condens. Matter \textbf{5}, 8763 (1993).

\bibitem{Vonsovskii_FMM_1993} S. V. Vonsovskii, M. I. Katsnelson, and A. V.
Trefilov, Fiz. Met. Metalloved. 76 (3) 3 (1993); 76 (4), 3 (1993).

\bibitem{2band-model} M. B. Stearns, Phys. Rev. B \textbf{8}, 4383 (1973);
R. Mota and M. D. Coutinho-Filho, Phys. Rev. B \textbf{33}, 7724 (1986).

\bibitem{Neugebauer} F. K\"ormann, A. Dick, B. Grabowski, B. Hallstedt, T.
Hickel, and J. Neugebauer, Phys. Rev. B \textbf{78}, 033102 (2008)

\bibitem{Gornostyrev} S.V. Okatov, Yu.N. Gornostyrev, A.I. Lichtenstein, and
M.I. Katsnelson, Phys. Rev. B \textbf{84}, 214422 (2011).

\bibitem{Lichtenstein_2001} A. I. Lichtenstein, M. I. Katsnelson, and G.
Kotliar, Phys. Rev. Lett. \textbf{87}, 067205 (2001).

\bibitem{Katanin2010} A. A. Katanin, A. I. Poteryaev, A. V. Efremov, A. O.
Shorikov, S. L. Skornyakov, M. A. Korotin, V. I. Anisimov, Phys. Rev. B 
\textbf{81}, 045117 (2010).

\bibitem{WernerQMC} P. Werner et al., Phys. Rev. Lett. \textbf{97}, 076405
(2006).

\bibitem{Abrikosov2013} L. V. Pourovskii, T. Miyake, S. I. Simak, A. V.
Ruban, L. Dubrovinsky, and I. A. Abrikosov, Phys. Rev. B \textbf{87}, 115130
(2013)

\bibitem{Barth_Hedin_1972} U. von Barth and L. Hedin, J. Phys. C 5, 1629
(1972).

\bibitem{Werner} P.~Werner, E.~Gull, M.~Troyer, and A.J.~Millis, Phys. Rev.
Lett. \textbf{101}, 166405 (2008).

\bibitem{Igoshev2013} P.A. Igoshev, A.V. Efremov, A.I. Poteryaev, A.A.
Katanin, V.I. Anisimov, Phys. Rev. B \textbf{88}, 155120 (2013).

\bibitem{Moriya} T. Moriya, Spin fluctuations in itinerant magnets.
Springer-Verlag, Berlin, Heidelberg, 1985.

\bibitem{Suppl} See Supplementary Material at http://

\bibitem{Schmalian} J\"{o}rg Schmalian, David Pines, and Branko Stojkovi\'{c}%
, Phys. Rev. B \textbf{60}, 667 (1999); A. Abanov, A. V. Chubukov, and J.
Schmalian, Adv. Phys. \textbf{52}, 119 (2003).

\bibitem{DGA1} A. A. Katanin, A. Toschi, and K. Held, Phys. Rev. B \textbf{80%
}, 075104 (2009).

\bibitem{Anisimov} V. I. Anisimov, A. S. Belozerov, A. I. Poteryaev, and I.
Leonov, Phys. Rev. B \textbf{86}, 035152 (2012).






\bibitem{Lichtenstein1987} A. I. Liechtenstein, M. I. Katsnelson, V. P.
Antropov, V. A. Gubanov, JMMM \textbf{67}, 65 (1987).










%





\bibitem{Mook1973} H.A.~Mook and R.M. Nicklow, Phys. Rev. B \textbf{7}, 336
(1973).

\bibitem{DGA} See, e.g. A. Toschi, A. A. Katanin, and K. Held, Phys. Rev. B 
\textbf{75}, 045118 (2007); A. Toschi, G. Rohringer, A. A. Katanin, K. Held,
Ann. der Phys., 523, \textbf{698} (2011).

\bibitem{DF} See, e.g., A. N. Rubtsov, M. I. Katsnelson, A. I. Lichtenstein,
A. Georges, Phys.Rev. B \textbf{79} 045133 (2009).



\end{thebibliography}

\begin{thebibliography}{}
\bibitem{Pourovskii} L.~V. Pourovskii, J.~Mravlje, M.~Ferrero, O.~Parcollet,
and I.A.~Abrikosov, Phys. Rev. B \textbf{90}, 155120 (2014).

\bibitem{OurJETP} P. A. Igoshev, A. A. Katanin, V. Yu. Irkhin, JETP \textbf{%
105}, 1043 (2007).
\end{thebibliography}
\end{document}